\documentclass[11pt]{article}

\usepackage{a4wide}
\usepackage{amsmath}
\usepackage{bm}
\usepackage{amssymb}
\usepackage{hyperref}
\usepackage{epsfig}
\usepackage{epic}
\usepackage{mathrsfs}
\usepackage[small,bf]{caption}
\usepackage[nosort]{cite}
\usepackage[bulletsep]{collref}

\newcommand{\beq}{\begin{equation}}
\newcommand{\eeq}{\end{equation}}

\newcommand{\beqa}{\begin{eqnarray}}
\newcommand{\eeqa}{\end{eqnarray}}

\makeatletter
\def\mr@ignsp#1 {\ifx\:#1\@empty\else #1\expandafter\mr@ignsp\fi}%
\newcommand{\multiref}[1]{\begingroup
\xdef\mr@no@sparg{\expandafter\mr@ignsp#1 \: }%
\def\mr@comma{}%
\@for\mr@refs:=\mr@no@sparg\do{\mr@comma\def\mr@comma{,}\ref{\mr@refs}}%
\endgroup}
\makeatother
\long\def\symbolfootnote[#1]#2{\begingroup%
\def\thefootnote{\fnsymbol{footnote}}\footnote[#1]{#2}\endgroup}

\numberwithin{equation}{section}
\begin{document}
\thispagestyle{empty}

\begin{flushright}\footnotesize
\texttt{UUITP-30/11}
\vspace{0.5cm}
\end{flushright}
\setcounter{footnote}{0}

\vspace{6mm}

\begin{center}
{\Large\textbf{\mathversion{bold} Generalised scaling at subleading order}}
\vspace{15mm}

{\sc Lisa Freyhult$^{a,}$\footnote{lisa.freyhult@physics.uu.se}}\\[10mm]

{\it $^a$  Department of Physics and Astronomy, Uppsala University\\
	    P.O. Box 803, S-75108, Uppsala, Sweden}\\[28mm]

\textbf{Abstract}\\[6mm]
\end{center}

\noindent{We study operators in the $sl(2)$ sector of $\mathcal{N}=4$ SYM in the generalised scaling limit, where the spin is large and the length of the operator scales with the logarithm of the spin. At leading order in the large spin expansion the scaling dimension at strong coupling is given in terms of the free energy of the $O(6)$ model. We investigate the first subleading corrections to the scaling dimension and find that these too can be derived from the $O(6)$ model in the strong coupling limit.
}
\newpage
\section{Introduction and summary}
In recent years much progress has been made in understanding the AdS/CFT correspondence and the theories related by it. Much of this progress is due to the unusual property of integrability for the constituent theories. This allows for the construction of a set of equations giving the spectrum of scaling dimensions in the planar theory. Originally the spectrum was constructed for long operators where the techniques of the asymptotic Bethe ansatz could be applied (for pedagogical reviews see \cite{Staudacher:2010jz,Ahn:2010ka,Vieira:2010kb}). This was later extended to include operators of arbitrary length and the spectrum in this case is given by a thermodynamic Bethe ansatz or a Y-system \cite{Bajnok:2010ke,Gromov:2010kf}. Recently, a proposal for a formulation of this system in terms of a finite number of non-linear integral equations (FiNLIE) was put forward \cite{Gromov:2011cx}.

In the construction of the solution to the spectral problem operators belonging to the $sl(2)$ sector of the theory play an important role. These operators are built from complex scalar fields and covariant light-cone derivatives and can schematically be written as
\begin{equation}\label{operator}
\mbox{Tr}(D^MZ^L)+\dots
\end{equation}
where $M$ is referred to as the spin of the operator and $L$ is the length or the twist. In the limit of large spin the operators exhibit logarithmic scaling \cite{Collins:1989bt,Korchemsky:1988si,Korchemsky:1992xv,Belitsky:2006en}, with an anomalous dimension
\begin{equation}\label{cusp}
\gamma(g,M,L)=f(g)\left(\log M+\gamma_E+(L-2)\log 2\right)+B_L(g)+\dots
\end{equation}
for the ground state.
The function $f(g)$ is the universal scaling function, or the cusp anomalous dimension, that can be constructed to all orders in $g$ using integrability \cite{Eden:2006rx,Beisert:2006ez}. The scaling function is a well tested quantity both at weak and strong coupling (see \cite{Freyhult:2010kc} for a review and references). Here we follow the convention where the coupling constant is related to the 't Hooft coupling as $g^2=\frac{\lambda}{16\pi^2}$. The new function appearing at subleading order, $B_L(g)$, was studied in \cite{Freyhult:2009my,Fioravanti:2009xt} and found to be in agreement with the corresponding quantity obtained from the string sigma model \cite{Beccaria:2008tg,Gromov:2011de}. Both $f(g)$ and $B_L(g)$ are quantities that can be tracked from weak to strong coupling. The interpolating properties of the functions make them good observables when studying the AdS/CFT correspondence. Considering the structure of the expansion in \eqref{cusp}, one could be led to assume that the expansion in $M$ would continue with new functions of the coupling constant appearing at each order. 
The dots in \eqref{cusp} are however quite non-trivial and interesting, and were studied for twist 2 operators in \cite{Gromov:2011de,Basso:2011rc}. The next order in the expansion is in that case of order $\tfrac{1}{\log M}$ at strong coupling while at weak coupling it is suppressed by $1/M$ \cite{Dokshitzer:2005bf,Dokshitzer:2006nm,Basso:2006nk,Beccaria:2010tb}. The strong coupling behavior was predicted from the string sigma model \cite{SchaferNameki:2005is,SchaferNameki:2006ey,Beccaria:2010ry,Giombi:2010zi} and reproduced by including finite size corrections in \cite{Gromov:2011de,Basso:2011rc}. It was however found that the large spin limit considered in the strong and weak coupling limits are two different scaling limits and in order to find a reconciliation between the two regimes the series has to be resummed \cite{Basso:2011rc}. 

Another interesting limit for the operators \eqref{operator} is the large spin limit where the twist scales logarithmically with the spin. This limit is referred to as generalised scaling. The anomalous dimension is in this case given by  \cite{Belitsky:2006en,Frolov:2006qe, Casteill:2007ct,Alday:2007mf,Freyhult:2007pz,Bombardelli:2008ah,Fioravanti:2008rv,Fioravanti:2008ak,Basso:2008tx,Bajnok:2008it}
\begin{equation}
\gamma(g,M,L)=\left(f(g)+\epsilon(g,\tfrac{L}{\log M})\right)\log M+\dots\,.
\end{equation}
At strong coupling, in the special limit where
\begin{equation}\label{limit}
g\to\infty,\quad M,\,L\to\infty,\quad\frac{L}{\log M}\frac{1}{m_{O(6)}}=\mbox{fixed},
\end{equation}
with $m_{O(6)}$ given by
\begin{equation}\label{mass}
m_{O(6)}=\frac{\sqrt2}{\Gamma(5/4)}(2\pi g)^{1/4}e^{-\pi g}\left(1+\mathcal{O}(1/g)\right),
\end{equation}
the full sigma model on $AdS_5\times S^5$ reduces to the $O(6)$ sigma model \cite{Alday:2007mf}. The scale, $m_{O(6)}$ coincides with the mass of the $O(6)$ model. In particular the anomalous dimension is given in terms of the free energy of the ground state of the $O(6)$ model, $\epsilon(g,\tfrac{L}{\log M})$. Here we will study the first subleading corrections in the generalised scaling limit. The subleading corrections were first considered in \cite{Fioravanti:2009xn} where the weak coupling expansion for small values of $\tfrac{L-2}{\log M}$ was worked out. The strong coupling expansion, in a corresponding limit, was also considered and the results showed an intriguing structure which suggested a possible relation to the $O(6)$ model. We revisit the calculation at strong coupling and explicitly demonstrate the relation between the subleading corrections and the $O(6)$ model. We believe that our analysis can be taken as a starting point for investigating further corrections, similar to the case of finite twist \cite{Basso:2011rc}, in the generalised scaling limit.

We find that the subleading corrections to the anomalous dimension in the limit\footnote{At leading order this coincides with the limit \eqref{limit}}
\begin{equation}\label{O(6)limit1}
g\to\infty,\quad L,\,M\to\infty,\quad \frac{L-2}{R(M,g)m_{O(6)}}=\mbox{fixed},
\end{equation}
with 
\begin{equation}
R(M,g)=\log \tfrac{2M}{g}+\dots,
\end{equation}
can be written in terms of the the free energy of the $O(6)$ model as well as the universal scaling function, $f(g)$, and the virtual scaling function, $B_L(g)$. The anomalous dimension reads
\begin{equation}
\gamma(g,L,M)=f(g)(\log M+\gamma_E+2\log 2)+B_L(g)+\epsilon(g,\tfrac{L-2}{R(M,g)})R(M,g)+\dots\,.
\end{equation}
We note that the strong coupling expansion is organised in terms of the natural expansion parameter of the string sigma model, $M/g$.

The paper is organised as follows. In the next section we recall the integral equation incorporating the leading, as well as the subleading corrections. Our approach is based on the results of \cite{Freyhult:2007pz}, the integral equation we derive is however completely equivalent to the integral equation constructed in \cite{Fioravanti:2009xn}. The integral equation is derived from the asymptotic Bethe ansatz for the $sl(2)$ sector where we construct a counting function so that the dynamics is described in terms of holes instead of Bethe roots (see \cite{Feverati:2006tg} for an introduction to the method and further references).

We proceed by rearranging the integral equation into a form suitable for analysis at strong coupling. This formulation is then used to analyse the equation for the hole density at strong coupling. Our result shows that this equation, in the particular limit \eqref{O(6)limit1}, is of the same form as the integral equation describing the ground state of the $O(6)$ model. Furthermore, the anomalous dimension is computed at strong coupling and found to be mapped to the free energy of the $O(6)$ model. Finally we identify the parameters of the model, which then allows us to write an explicit expression for the anomalous dimension in the relevant limit.

\section{An integral equation for subleading corrections}
A particularly useful way to organise the large spin expansion of twist two operators in the $sl(2)$ sector is to use the counting function. This approach was introduced in \cite{Destri:1992qk,Destri:1994bv,Fioravanti:1996rz} and widely used therafter (see \cite{Feverati:2006tg} and references therein). Studying the counting function instead of the density of Bethe roots amounts to a change of reference state, where the dynamical variables are holes rather than Bethe roots. For the $sl(2)$ sector the number of holes equals the length of the operator, and hence the twist two case becomes particularly simple \cite{Belitsky:2006en,Freyhult:2007pz,Bombardelli:2008ah}.

In the following we will be interested in large spin operators in the $sl(2)$ sector where the length scales with the logarithm of the spin. In this case the advantage of the method is less obvious, nevertheless we will demonstrate in the following the usefulness of constructing the solution in terms of the hole density.

Restricting to the ground state, the counting function, $Z(u)$, becomes an odd function. Keeping only the terms relevant for the leading and first subleading corrections, it reads \cite{Freyhult:2007pz} for $t>0$
\begin{eqnarray}\label{counting}
\nonumber\hspace{-0.8cm}\lefteqn{t\hat{Z}(t)=\frac{2\pi Le^{t/2}J_0(2gt)}{i(e^t-1)}-\frac{4\pi\cos(\tfrac{Mt}{\sqrt2})}{i(e^t-1)}-\frac{4\pi}{i}\frac{t}{e^t-1}\log 2\,\delta(t)-\frac{2\pi}{i(e^t-1)}\int_{-a}^adu\rho_h(u)\cos(tu)}\\
\hspace{-0.8cm}&-&\frac{4g^2te^{t/2}}{e^t-1}\int_0^\infty dt'e^{-t'/2}K(2gt,2gt')\left(t'\hat{Z}(t')-\frac{4\pi}{i}\cos(\tfrac{Mt'}{\sqrt2})-\frac{2\pi}{i}\int_{-a}^adu\rho_h(u)\cos(t' u)\right)
\end{eqnarray}
where we have anticipated logarithmic scaling, $L\sim \log M$.
In total there are $L$ holes. Two of them are large, of order $M$ and are responsible for the explicit $M$ dependence in \eqref{counting}. The remaining $L-2$ holes are represented by means of the density, $\rho_h(u)$, defined as
\begin{equation}\label{rhoholes}
2\pi\rho_h(u)=\int_0^\infty \frac{dt}{2\pi}\hat{Z}(t) t\cos(tu)2i,\quad -a\leq u\leq a,\quad\mbox{with}\quad \int_{-a}^a\rho_h(u)du=L-2.
\end{equation}
In Fourier space we have
\begin{equation}
\hat{\rho}_h(t)=2i\int_0^\infty \frac{dt'}{\pi}K_h(t,t')t'\hat{Z}(t').
\end{equation}
with
\begin{equation}
K_h(t,t')=\int_{-a}^a\frac{du}{4\pi}\cos(u t)\cos(ut').
\end{equation}
The kernel in \eqref{counting} is the BES kernel \cite{Beisert:2006ez} which can be represented as
\begin{equation}\label{BESkernel1}
K(2gt,2gt')=K_+(2gt,2gt')+K_-(2gt,2gt')+8g^2\int_0^\infty dt''\frac{t''}{e^{t''}-1}K_-(2gt,2gt'')K_+(2gt'',2gt')
\end{equation}
with
\begin{eqnarray}\label{BESkernel2}
\nonumber K_+(t,t')&=&\frac{2}{tt'}\sum_{n=1}^\infty (2n-1)J_{2n-1}(t)J_{2n-1}(t')\\
 K_-(t,t')&=&\frac{2}{tt'}\sum_{n=1}^\infty 2n J_{2n}(t)J_{2n}(t').
\end{eqnarray}

Note that if we had kept $L$ finite while expanding for large $M$, the endpoints given by $a$ would have vanished at this order and the $L-2$ holes would have been concentrated around the origin. In this case the counting function reduces to the one for the so called virtual corrections \cite{Freyhult:2009my, Fioravanti:2009xt}.

Substituting \eqref{rhoholes} into \eqref{counting} we obtain the integral equation for the counting function,
\begin{eqnarray}
\lefteqn{t\hat{Z}(t)=\frac{2\pi Le^{t/2}J_0(2gt)}{i(e^t-1)}-\frac{4\pi\cos(\tfrac{Mt}{\sqrt2})}{i(e^t-1)}-\frac{4\pi}{i}\frac{t}{e^t-1}\log2\,\delta(t)}\\
&-&\frac{4te^{t/2}}{(e^t-1)}\int_0^\infty dt' e^{-t'/2}\tilde{\mathcal{K}}(t,t')t'\hat{Z}(t')
\nonumber+\frac{16\pi te^{t/2}}{i(e^t-1)}\int_0^\infty dt'g^2e^{-t'/2}K(2gt,2gt')\cos(\tfrac{Mt'}{\sqrt2}).
\end{eqnarray}
To simplify notation, we have introduced the kernel
\begin{equation}
\tilde{\mathcal{K}}(t,t')=\frac{e^{-t/2}e^{t'/2}}{t}K_h(t,t')+g^2K(2gt,2gt')-4g^2\int_0^\infty dt''e^{-t''/2}e^{t'/2}K(2gt,2gt'')K_h(t'',t').
\end{equation}
To facilitate the large $M$ expansion we introduce
\begin{equation}\label{sigmadef}
\hat{\sigma}(t)=-\frac{i}{16\pi}e^{-t/2}\left(t\hat{Z}(t)+\frac{4\pi}{i}\frac{\cos(\tfrac{Mt}{\sqrt2})+\log 2\,t\,\delta(t)-1}{e^t-1}\right).
\end{equation}
After expanding for large values of $M$ and keeping leading and subleading terms we find 
\begin{eqnarray}\label{sigma}
\nonumber\hat{\sigma}(t)&=&\frac{t}{e^t-1}\left(-\frac{L}{8}\frac{J_0(2gt)}{t}+\frac{e^{-t/2}}{4t}+\tilde{\mathcal{K}}(t,0)\left(\log M+\gamma_E+2\log 2\right)\right.\\
&+&\left.\int_0^\infty dt'\left(\tilde{\mathcal{K}}(t,t')-\tilde{\mathcal{K}}(t,0)e^{t'}\right)\frac{e^{-t'/2}}{e^{t'}-1}-4\int_0^\infty dt'\tilde{\mathcal{K}}(t,t')\hat{\sigma}(t')\right).
\end{eqnarray}
This integral equation together with the anomalous dimension expressed in terms of $\hat{\sigma}(t)$ determines the leading and subleading corrections to the anomalous dimension in the large spin expansion. For technical reasons, and in particular to stress the relationship with the $O(6)$ model, it will prove convenient to instead work with the corresponding expression for the hole density \eqref{rhoholes},
\begin{eqnarray}\label{holes}
\nonumber\frac{1}{8}\hat{\rho}_h(t)&=&K_h(t,0)\left(\log M+\gamma_E+2\log 2\right)+\int_0^\infty dt'(K_h(t,t')-K_h(t,0)e^{t'/2})\frac{1}{e^{t'}-1}\\
&-&4\int_0^\infty dt'K_h(t,t')e^{t'/2}\hat{\sigma}(t'),
\end{eqnarray}
as was done for the leading order in \cite{Basso:2008tx}.
Further, we split $\hat{\sigma}(t)$ as follows
\begin{equation}\label{sigmasplitF}
\hat{\sigma}(t)=\frac{1}{e^t-1}\left(\frac{g}{2}\gamma(2gt)-\frac{L}{8}{J_0(2gt)}+\frac{e^{-t/2}}{4}+\frac{e^{-t/2}\hat{\rho}_h(t)}{8}\right).
\end{equation}
This allows us to rewrite \eqref{sigma} into
\begin{eqnarray}\label{gamma}
\nonumber \gamma(2gt)&=&\frac{2t}{g}\left[{\mathcal{K}}(t,0)\left(\log M+\gamma_E+2\log 2\right)+\int_0^\infty dt'\left({\mathcal{K}}(t,t')-{\mathcal{K}}(t,0)e^{t'}\right)\frac{e^{-t'/2}}{e^{t'}-1}\right.\\
&-&4\left.\int_0^\infty dt'\mathcal{K}(t,t')\hat{\sigma}(t')\right]
\end{eqnarray}
with the kernel
\begin{equation}
{\mathcal{K}}(t,t')=g^2K(2gt,2gt')-4g^2\int_0^\infty dt''e^{-t''/2}e^{t'/2}K(2gt,2gt'')K_h(t'',t').
\end{equation}
The definition \eqref{sigmadef} allows for a simple expression for the anomalous dimension,
\begin{equation}\label{AD1}
\gamma=16g^2\lim_{t\to0}\frac{\gamma(2gt)}{2gt}.
\end{equation}
To the relevant order in the large spin expansion the normalisation condition \eqref{rhoholes} reduce to 
\begin{eqnarray}\label{norm}
&&\nonumber\frac{4a}{\pi}(\log M+\gamma_E+2\log 2)+8\int_0^\infty dt(K_h(t,0)-K_h(0,0))\frac{1}{e^t-1}\\
&&-32\int_0^\infty dt K_h(t,0)e^{t/2}\hat{\sigma}(t)=\hat{\rho}_h(0)=L-2.
\end{eqnarray}
The integral equations presented above, taken together with the anomalous dimension \eqref{AD1} and the normalisation condition \eqref{norm}, are completely equivalent to the equations in \cite{Fioravanti:2009xn}.

\section{Strong coupling analysis}
Here we will analyse the subleading corrections by applying the methods developed for the leading order \cite{Basso:2008tx, Basso:2009gh}.
The first step is to write the equation \eqref{gamma} on a suitable form for studying strong coupling. This will then be used to establish that the strong coupling expansion of the equation for the hole density coincides, in form, with the integral equation for the ground state of the $O(6)$ model. Finally we use Wronskian relations similarly to \cite{Basso:2008tx,Freyhult:2009my, Basso:2009gh} to relate the parameters in the models.

 To rewrite \eqref{gamma} on a suitable for for the large $g$ expansion we follow the approach of splitting $\gamma(t)$ and the equations into an even and an odd part \cite{Basso:2007wd, Eden},
\begin{equation}\label{gammasplit}
\gamma(t)=\gamma_+(t)+\gamma_-(t).
\end{equation}
Using the notation for the kernels in \eqref{BESkernel1}-\eqref{BESkernel2} we obtain,
\begin{eqnarray}
\hspace{-0.1cm}\nonumber\frac{\gamma_+(2gt)}{2gt}&=&K_-(2gt,0)\left(\log M\hspace{-0.1cm}+\hspace{-0.1cm}\gamma_E\hspace{-0.1cm}+\hspace{-0.1cm}2\log 2\right)+\hspace{-0.1cm}\int_0^\infty\hspace{-0.1cm} dt'\left(K_-(2gt,2gt')-K_-(2gt,0)e^{t'}\right)\frac{e^{-t'/2}}{e^{t'}-1}\\
&-&\nonumber4\int_0^\infty dt'K_-(2gt,2gt')\left(\hat{\sigma}(t')+\frac{1}{8}e^{-t'/2}\rho_h(t')-\frac{g}{e^{t'}-1}\gamma_-(2gt')\right).\\
\hspace{-0.1cm}\nonumber\frac{\gamma_-(2gt)}{2gt}&=&K_+(2gt,0)\left(\log M\hspace{-0.1cm}+\hspace{-0.1cm}\gamma_E\hspace{-0.1cm}+\hspace{-0.1cm}2\log 2\right)+\hspace{-0.1cm}\int_0^\infty\hspace{-0.1cm} dt'\left(K_+(2gt,2gt')-K_+(2gt,0)e^{t'}\right)\frac{e^{-t'/2}}{e^{t'}-1}\\
&-&4\int_0^\infty dt'K_+(2gt,2gt')\left(\hat{\sigma}(t')+\frac{1}{8}e^{-t'/2}\rho_h(t')\right)
\end{eqnarray}
Expanding the even and the odd part of $\gamma(t)$ in a Neumann series \cite{Basso:2007wd},
\begin{eqnarray}\label{Neumann}
\nonumber\gamma_-(t)&=&2\sum_{n=1}^\infty (2n-1)J_{2n-1}(t)\gamma_{2n-1}\\
\gamma_+(t)&=&2\sum_{n=1}^\infty 2n J_{2n}(t)\gamma_{2n},
\end{eqnarray}
the equations can be expressed as
\begin{eqnarray}\label{strongcouplingform}
&&\nonumber\int_0^\infty \frac{dt}{t}J_{2n}(2gt)\left(\frac{\gamma_+(2gt)}{1-e^{-t}}-\frac{\gamma_-(2gt)}{e^t-1}\right)=h_{2n}\\
&&\int_0^\infty \frac{dt}{t}J_{2n-1}(2gt)\left(\frac{\gamma_+(2gt)}{e^t-1}+\frac{\gamma_-(2gt)}{1-e^{-t}}\right)=h_{2n-1}+\frac{1}{2}\delta_{n,1}(\log M+\gamma_E+2\log 2)
\end{eqnarray}
with
\begin{eqnarray}
 h_n&=&\frac{1}{2}\int_0^\infty dt\frac{J_n(2gt)}{2gt}\left(L\frac{J_0(2gt)}{e^t-1}-\frac{e^{t/2}\hat{\rho}_h(t)}{e^t-1}\right)-\frac{\delta_{n,1}}{2}\int_0^\infty dt\frac{e^{t/2}}{e^t-1}.
\end{eqnarray}
After a change of variables \cite{Basso:2007wd},
\begin{equation}\label{changeofvar}
\gamma_+(it)+i\gamma_-(it)=\frac{\sin(\tfrac{t}{4g})}{\sqrt2\sin(\tfrac{t}{4g}+\tfrac{\pi}{4})}\left(\Gamma_+(it)+i\Gamma_-(it)\right),
\end{equation}
and by making use of the Jacobi-Anger expansion, \eqref{strongcouplingform} gives
\begin{eqnarray}\label{JA}
&&\nonumber\int_0^\infty dt\left(e^{itu}\Gamma_-(t)-e^{-itu}\Gamma_+(t)\right)=2(\log M+\gamma_E+2\log 2)\\
&&+\int_0^\infty dt\left(\left(L\frac{J_0(2gt)}{e^t-1}
-\frac{e^{t/2}}{e^t-1}\hat{\rho}_h(t)\right)e^{2igtu}-2\frac{e^{t/2}}{e^t-1}\right), \quad -1\leq u\leq1.
\end{eqnarray}

We stress that we have merely rearranged the equations in order to facilitate the strong coupling analysis. The above equations could also be used to study the weak coupling limit, though this would be less convenient. 
\subsection{The hole density at strong coupling}
In the following we will study the strong coupling limit of the hole density. The starting point is the expression for the density, obtained by the inverse Fourier transform of \eqref{holes} where $\hat{\sigma}(t)$ is exchanged for $\gamma(t)$ by means of \eqref{sigmasplitF},
\begin{eqnarray}\label{holeeq}
\hspace{-0.5cm}\nonumber\lefteqn{\frac{1}{8}\int_{-\infty}^\infty \frac{dt}{2\pi}e^{ikt}\hat{\rho}_h(t)=\frac{1}{4\pi}(\log M+\gamma_E+2\log 2)+\int_0^\infty \frac{dt}{4\pi}(\cos(kt)-e^{t/2})\frac{1}{e^t-1}}\\
\hspace{-0.5cm}&-&4\int_0^\infty\frac{dt}{4\pi}\cos(kt)\frac{e^{t/2}}{e^t-1}\left(\frac{g}{2}\gamma(2gt)-\frac{L}{8}J_0(2gt)+\frac{e^{-t/2}}{4}+\frac{e^{-t/2}}{8}\hat{\rho}_h(t)\right),\quad  -a\leq k\leq a
\end{eqnarray}
Equation \eqref{holeeq} contains the integral 
\begin{equation}\label{integral}
I(k)=-g\int_0^\infty \frac{dt}{2\pi}\cos(kt)\frac{e^{t/2}}{e^t-1}\gamma(2gt),
\end{equation}
which we rewrite as in \cite{Basso:2008tx}. We change variables as in \eqref{changeofvar}, use the relations,
\begin{eqnarray}
&&\cos(kt)\frac{\cosh(\tfrac{t}{4g})}{\cosh(\tfrac{t}{2g})}=\sqrt{2}g\int_{-\infty}^\infty du\cos(ut)\frac{\cosh(g\pi(u+k))}{\cosh(2g\pi(u+k))}\\
&&\cos(kt)\frac{\sinh(\tfrac{t}{4g})}{\cosh(\tfrac{t}{2g})}=\sqrt{2}g\int_{-\infty}^\infty du\sin(ut)\frac{\sinh(g\pi(u+k))}{\cosh(2g\pi(u+k))},
\end{eqnarray}
and find
\begin{eqnarray}\label{integralnewvar}
\nonumber \pi I(k)=-\frac{g}{4\sqrt2}\int_0^\infty dt\int_{-\infty}^\infty du&&\Big(\cos(ut)\frac{\cosh(g\pi(u+\tfrac{k}{2g}))}{\cosh(2g\pi(u+\tfrac{k}{2g}))}(\Gamma_-(t)-\Gamma_+(t))\\
&&+\sin(ut)\frac{\sinh(g\pi(u+\tfrac{k}{2g}))}{\cosh(2g\pi(u+\tfrac{k}{2g}))}(\Gamma_-(t)+\Gamma_+(t))\Big).
\end{eqnarray}
Splitting the integral over $u$ as $\int_{-\infty}^\infty=\int_{-1}^1+\int_1^\infty+\int_{-\infty}^{-1}$ \eqref{JA} can be used and the large $g$ limit taken (see Appendix A for details).We then find
\begin{eqnarray}\label{integralres}
\nonumber I(k)&=&-\frac{1}{4\pi}\left(\log M+\gamma_E+2\log 2\right)+\frac{\tilde{m}}{16}\cosh\left(\tfrac{\pi k}{2}\right)\\
&-&\frac{1}{4\pi}\int_0^\infty dt\left[\left(L\frac{J_0(2gt)}{e^t-1}-\frac{e^{t/2}\hat{\rho}_h(t)}{e^t-1}\right)\frac{e^{t/2}}{e^t+e^{-t}}\cos(kt)-\frac{e^{t/2}}{e^t-1}\right]
\end{eqnarray}
where
\begin{eqnarray}\label{mass1}
\nonumber \lefteqn{\frac{\tilde{m}}{16}=\frac{1}{\sqrt2\pi^2}e^{-\pi g}\left(\log M+\gamma_E+2\log 2\right)-\frac{g}{2\pi}e^{-\pi g}\int_0^\infty dt\,\mbox{Re}\left[\frac{\Gamma_++i\Gamma_-}{t+i\pi g}e^{i(t-\pi/4)}\right]}\\
&+&\frac{1}{4\pi}e^{-\pi g}\int_0^\infty dt\left[\left(L\frac{J_0(2gt)}{e^t-1}-\frac{e^{t/2}\hat{\rho}_h(t)}{e^t-1}\right)\mbox{Re}\left[\frac{i}{t+i\pi/2}e^{i(2gt-\pi/4)}\right]-\frac{2\sqrt2}{\pi}\frac{e^{t/2}}{e^t-1}\right].
\end{eqnarray}
With the result of \eqref{integralres}, \eqref{holeeq} becomes
\begin{eqnarray}
\nonumber\int_{-\infty}^\infty\frac{dt}{2\pi} \hat{\rho}_h(t)e^{ikt}&=&\tilde{m}\cosh(\tfrac{\pi k}{2})+\frac{L}{\pi}\int_0^\infty dtJ_0(2gt)\cos(kt)\frac{\sinh(\frac{t}{2})}{\cosh(t)}\\
&+&\frac{1}{\pi}\int_0^\infty dt\cos(kt)\frac{e^t+1}{e^{2t}+1}\hat{\rho}_h(t).
\end{eqnarray}
The integral containing the Bessel function, $J_0(2gt)$, can be computed at strong coupling \cite{Basso:2008tx},
\begin{equation}
\frac{L}{\pi}\int_0^\infty dt J_0(2gt)\cos(kt)\frac{\sinh(\tfrac{t}{2})}{\cosh(t)}=\frac{L}{\pi\sqrt g}e^{-\pi g}\cosh\tfrac{\pi k}{2}\left(1+\mathcal{O}(1/g)\right)
\end{equation}

We conclude that the holes satisfy the following equation, to leading order in the large $g$ expansion,
\begin{eqnarray}\label{eq:rhohole}
&&\rho_h(\theta)=m\cosh(\theta)+\int_{-\tfrac{\pi a}{2}}^{\tfrac{\pi a}{2}}d\theta'\rho_h(\theta')K(\theta-\theta')\\
&&\nonumber\mbox{with}\quad K(\theta)=\frac{1}{4\pi^2}\left(\psi(1+\tfrac{i}{2\pi}\theta)+\psi(1-\tfrac{i}{2\pi}\theta)-\psi(\tfrac{1}{2}+\tfrac{i}{2\pi}\theta)-\psi(\tfrac{1}{2}-\tfrac{i}{2\pi}\theta)+\frac{2\pi}{\cosh(\theta)}\right).
\end{eqnarray}
In the above we  have introduced
\begin{equation}
m=\tilde{m}+\frac{L}{\pi\sqrt g}e^{-\pi g}+\dots
\end{equation}
and $\theta=\pi k/2$. This integral equation is of the same form as the integral equation for the ground state of the $O(6)$ model \cite{Zamolodchikov:1977nu,Hasenfratz:1990zz,Hasenfratz:1990ab}. In the next section we will discuss how the parameters of the models are related.

It is possible to express the anomalous dimension at strong coupling in terms of known functions and the hole density as follows (we refer to Appendix B for details),
\begin{eqnarray}\label{AD}
\gamma&=&16g^2\gamma_1^{(0)}\left(\log M+\gamma_E+2\log 2\right)+16g^2\gamma_1^{(1)}\\
\nonumber \gamma_1^{(1)}
&=&\frac{m_{O(6)}}{8g^2}\int_{-a}^a dk\rho_h(k)\left(\cosh(\tfrac{\pi k}{2})-1\right)+\frac{1}{16g^2}B_L(g)
\end{eqnarray}
where $m_{O(6)}$ is the mass of the $O(6)$ model \eqref{mass}.

\subsection{Relating the parameters}
We have seen that the structure of the subleading corrections show a similarity to the $O(6)$ model. In order to quantify this similarity we compute the parameter $m$.

The parameter $m$ in the effective equation for the holes contains the functions $\Gamma_\pm(t)$ which are solutions to the system of equations \eqref{strongcouplingform}. To leading order in the large $M$ limit these equations were solved in \cite{Basso:2007wd,Basso:2009gh}. The analysis for equations of this type is quite involved, but luckily in this case there is no need to explicitly solve the equations. We will be able to write the expression for the parameter $m$ in terms of known quantities. The expression for $m$ will involve the leading large $M$ solution to the two first orders in the expansion of the non-perturbative parameter $m_{O(6)}$.

In \cite{Basso:2009gh} the non-perturbative corrections to the scaling function, $f(g)$, were analysed and the relevant corrections to the density computed. This involved the study of the equations 
\begin{eqnarray}\label{nonpert}
&&\nonumber\int_0^\infty\frac{dt}{t}J_{2n}(t)\left(\frac{\hat{\gamma}_+(t)}{1-e^{-t/2g}}-\frac{\hat{\gamma}_-(t)}{e^{t/2g}-1}\right)=\hat{h}_{2n}(g)\\
&&\int_0^\infty\frac{dt}{t}J_{2n-1}(t)\left(\frac{\hat{\gamma}_-(t)}{1-e^{-t/2g}}+\frac{\hat{\gamma}_+(t)}{e^{t/2g}-1}\right)=\hat{h}_{2n-1}(g)
\end{eqnarray}
where
\begin{eqnarray}
\nonumber&&\hat{h}_{2n}=\frac{2m'}{\pi}\int_0^\infty dt\frac{J_{2n}(t)}{t^2+(\pi g)^2}\left(\frac{\pi g}{e^{t/2g}-1}+\frac{t}{1-e^{-t/2g}}\right)\\
 &&\hat{h}_{2n-1}=\frac{2m'}{\pi}\int_0^\infty dt\frac{J_{2n-1}(t)}{t^2+(\pi g)^2}\left(\frac{t}{e^{t/2g}-1}-\frac{\pi g}{1-e^{-t/2g}}\right).
\end{eqnarray}
The parameter $m'$ was introduced in \cite{Basso:2009gh} and is proportional to $m_{O(6)}$. We will keep the factors $m'$ in place to conform to the same notation as in that paper. Our final result will however not depend on the parameter $m'$. 
We will make use of the fact that the solution to these equations is known. The form of the left hand side of \eqref{nonpert} coincides with our equations \eqref{strongcouplingform}, while the right hand sides differ. This motivates the introduction of an auxiliary variable, $x$, and the construction of the following system of equations,
\begin{eqnarray}
\nonumber&&\int_0^\infty\frac{dt}{t}J_{2n}(t)\left(\frac{{\gamma}_+(t,x)}{1-e^{-t/2g}}-\frac{{\gamma}_-(t,x)}{e^{t/2g}-1}\right)=(1-x)h_{2n}+x\hat{h}_{2n}(g)\\
\nonumber&&\int_0^\infty\frac{dt}{t}J_{2n-1}(t)\left(\frac{{\gamma}_-(t,x)}{1-e^{-t/2g}}+\frac{{\gamma}_+(t,x)}{e^{t/2g}-1}\right)=\frac{1}{2}(1-x)\delta_{n,1}(\log M+\gamma_E+2\log 2)\\
&&\hspace{7cm}+\,(1-x)h_{2n-1}+x\,\hat{h}_{2n-1}(g).
\end{eqnarray}
$x=1$ corresponds to the system \eqref{nonpert} and $x=0$ corresponds to \eqref{strongcouplingform}, the equations for $\gamma(t)$.

Multiplying the equations by $2(2n)\gamma_{2n}(x')$ and $2(2n-1)\gamma_{2n-1}(x')$ respectively, summing over $n$ and finally subtracting the two equations from each other leads to the relation \cite{Basso:2008tx}
\begin{eqnarray}
\hspace{-0.5cm}&&\nonumber\sum_n\left((1-x)h_{2n}2(2n)\gamma_{2n}(x')+x\hat{h}_{2n}2(2n)\gamma_{2n}(x')\right)-\gamma_1(x')(1-x)(\log M+\gamma_E+2\log 2)\\
\hspace{-0.5cm}&-&\sum_n\left((1-x)h_{2n-1}2(2n-1)\gamma_{2n-1}(x')+x\hat{h}_{2n-1}2(2n-1)\gamma_{2n-1}(x')\right)-(x\leftrightarrow x')=0.
\end{eqnarray}
Setting $x=1$ and $x'=0$ gives
\begin{eqnarray}\label{Wronskianresult}
&&\nonumber\sum_n\big(\,\hat{h}_{2n-1}2(2n-1)\gamma_{2n-1}-\hat{h}_{2n}2(2n)\gamma_{2n}\,\big)\\
&&=\hat{\gamma}_1(\log M+\gamma_E+2\log 2)+\sum_n\big(\,{h}_{2n-1}2(2n-1)\hat{\gamma}_{2n-1}-{h}_{2n}2(2n)\hat{\gamma}_{2n}\,\big).
\end{eqnarray}
Evaluating the right and left hand sides and keeping terms to leading order in the non-perturbative expansion $\mathcal{O}(e^{-\pi g})$ we find after some algebra (see Appendix C)
\begin{eqnarray}\label{Wronskianlargeg}
&&\hspace{-0.9cm}\nonumber-\frac{8g}{\pi}e^{-\pi g}\int_0^\infty \mbox{Re}\left(\frac{\Gamma_+(t)+i\Gamma_-(t)}{t+i\pi g}e^{i(t-\pi/4)}\right)\\
&&\hspace{-0.9cm} \nonumber =\left(m_{O(6)}-\frac{8\sqrt2}{\pi^2}e^{-\pi g}\right)(\log M+\gamma_E+2\log 2)\\
&& \hspace{-0.9cm}\nonumber-\frac{4}{\pi}e^{-\pi g}\int_0^\infty dt\left[\left(L\frac{J_0(2gt)}{e^t-1}-\frac{e^{t/2}}{e^t-1}\hat{\rho}_h(t)\right)\mbox{Re}\left(\frac{ie^{-i\pi/4}e^{2igt}}{t+i\pi/2}\right)-\frac{2\sqrt2}{\pi}\frac{e^{t/2}}{e^t-1}\right]\\
&&\hspace{-0.9cm}+\frac{2\sqrt2 g}{m'}\int_0^\infty dt\left[\left(L\frac{J_0(2gt)}{e^t-1}-\frac{e^{t/2}}{e^t-1}\hat{\rho}_h(t)\right)\frac{\delta\gamma_-(2gt)-\delta\gamma_+(2gt)}{4gt}-\frac{m' m_{O(6)}}{2\sqrt{2}g}\frac{e^{t/2}}{e^t-1}\right].
\end{eqnarray}
Substituting this expression into the mass parameter \eqref{mass1} we find that $m$ can be expressed entirely in terms of known functions,
\begin{eqnarray}\label{mass2}
 \hspace{-0.5cm}\lefteqn{m=m_{O(6)}(\log M+\gamma_E+2\log 2)+\frac{L}{\pi\sqrt{g}}e^{-\pi g}}\\
\nonumber\hspace{-0.5cm}&+&\frac{2\sqrt2 g}{m'}\int_0^\infty dt\left[\left(L\frac{J_0(2gt)}{e^t-1}-\frac{e^{t/2}}{e^t-1}\hat{\rho}_h(t)\right)\frac{\delta\gamma_-(2gt)-\delta\gamma_+(2gt)}{2gt}-\frac{\sqrt2 m_{O(6)}m'}{4g}\frac{e^{t/2}}
{e^t-1}\right].
\end{eqnarray}
In the above we have used the notation introduced in \cite{Basso:2009gh},
\begin{equation}
\hat{\gamma}_-(2gt)-\hat{\gamma}_+(2gt)=\delta\gamma_-(2gt)-\delta\gamma_+(2gt)+\frac{2\sqrt2m_{O(6)}t}{\pi}\mbox{Re}\left(\frac{ie^{-i\pi/4}}{t+i\pi/2}\right).
\end{equation}

$\delta\gamma_\pm$ scales with $m_{O(6)}^2$. Comparing with the $O(6)$ model and the limit in which it appears at leading order we conclude that the scaling when including subleading corrections is $L-2\sim m_{O(6)}\log M$ (see also \cite{Fioravanti:2009xn}). Hence the terms in \eqref{mass2} proportional to $L-2$ will be subleading at large coupling. Further, the hole density is proportional to $m_{O(6)}$ and the terms in \eqref{mass2} containing it will be subleading as well. 
With the change of variables, analogous to \eqref{changeofvar} \cite{Basso:2009gh},
\begin{equation}
\delta\gamma_+(it)+i\delta\gamma_-(it)=\frac{\sin(\tfrac{t}{4g})}{\sqrt2\sin(\tfrac{t}{4g}+\tfrac{\pi}{4})}\left(\delta\Gamma_+(it)+i\delta\Gamma_-(it)\right),
\end{equation}
we hence find that the terms surviving in the scaling limit are
\begin{eqnarray}\label{mass5}
m&=&\nonumber m_{O(6)}\left(\log M+\gamma_E+2\log 2\right)+\frac{2}{\pi \sqrt{g}}e^{-\pi g}\\
&-&\frac{4\sqrt2 g}{m'}\int_0^\infty\left(\frac{\delta\Gamma_-(2gt)+\delta\Gamma_+(2gt)}{2gt}+\frac{\sqrt{2}m_{O(6)}m'}{8g}\frac{e^{t/2}}{e^t-1}\right).
\end{eqnarray}
This is evaluated using the solution for $\delta\Gamma_\pm(t)$ \cite{Basso:2009gh} (see appendix C) and we find 
\begin{equation}\label{mass4}
m=m_{O(6)}\left(\log \tfrac{2M}{g}+\dots\right).
\end{equation}

\subsection{The anomalous dimension at strong coupling}
From the above considerations we conclude that in the limit
\begin{equation}\label{O(6)limit}
g\to\infty,\quad L,\,M\to\infty,\quad \frac{L-2}{R(M,g)m_{O(6)}}=\mbox{fixed}
\end{equation}
with 
\begin{equation}
R(M,g)=\log \tfrac{2M}{g}+\dots,
\end{equation}
the anomalous dimension can be expressed in terms of the hole density \eqref{AD}. The hole density satisfies the integral equation of the $O(6)$ model \eqref{eq:rhohole} with the mass parameter \eqref{mass4}. Further the density is normalised as \eqref{rhoholes}. From this we conclude that the anomalous dimension can be written in terms of the free energy of the ground state of the $O(6)$ model, $\epsilon(g,\tfrac{L-2}{R(M,g)})$, and the functions, $f(g)$ and $B_L(g)$, that appear at leading order
\begin{equation}
\gamma(g,L,M)=f(g)(\log M+\gamma_E+2\log 2)+B_L(g)+\epsilon(g,\tfrac{L-2}{R(M,g)})R(M,g).
\end{equation}

Keeping the first terms at strong coupling the anomalous dimension becomes
\begin{equation}\label{final}
\gamma(g,L,M)=\left(f(g)+\epsilon\left(g,(L-2)/\log\tfrac{M}{g}\right)\right)\log \frac{M}{g}+\dots\,.
\end{equation}
Our analysis of the strong coupling expansion as well as the result \eqref{final} differs from \cite{Fioravanti:2009xn}. In \cite{Fioravanti:2009xn} the equations were first expanded for small values of $a$ and then evaluated at small and large coupling. We believe that the expansion for small $a$ is problematic for the subleading corrections at strong coupling. Our results are in line with what should be expected from string theory \cite{Alday:2007mf} where the natural expansion parameter is large $M/g$, in contrast to large $M$ as in the gauge theory.
\section{Conclusions}
We studied the subleading corrections to the anomalous dimension in the generalised scaling limit. At strong coupling we found that the integral equations derived from the asymptotic Bethe ansatz reduce to the equations for the ground state of the $O(6)$ model. This was observed to happen at leading order in the large spin expansion and here we conclude that the behavior generalise also to the subleading order. Further, the anomalous dimension was written in terms of the free energy of the $O(6)$ model. This fact is particularly convenient as it allows us to write down explicit expressions for the scaling dimension in various limits by exploiting results already existing in the literature \cite{Hasenfratz:1990zz, Hasenfratz:1990ab,Basso:2008tx,Bajnok:2008it,Volin:2009wr}.

Our results seem quite natural from the sigma model point of view \cite{Alday:2007mf}  since the natural expansion parameter in that case is not large $M$ but rather large $M/g$. It would however be interesting to understand the subleading corrections  in more detail from the sigma model.

It would also be interesting to continue the large spin expansion to the next order, $\mathcal{O}(\tfrac{1}{\log M})$. Here we expect the relation to the $O(6)$ model to break down. It would be interesting to study the corrections to it, in particular since at this order finite size corrections should start contributing \cite{Basso:2011rc}.

The $O(6)$ model appears when studying the lowest energy state of the spinning folded string or the corresponding operator. The highest excited state corresponds to the spiky string \cite{Kruczenski:2004wg} and the structure of the Bethe ansatz solution to leading order is quite similar to the lowest energy state \cite{Freyhult:2009bx}. In this case we would expect the relation to the $O(6)$ model to be broken at first subleading order but it could be interesting to study this in more detail.

\subsection*{Acknowledgements}
I would like to thank J. Minahan, A. Rej and S. Zieme for discussions and helpful suggestions. I also thank J. Minahan for reading the manuscript and B. Basso, D. Fioravanti and M. Rossi for helpful correspondence. This research was supported in part by the Swedish research council (VR).
\appendix
\section{Computing the integral $I(k)$}
We consider the evaluation of the integral \eqref{integralnewvar} at large coupling.
We split the integral over $u$ as, $\int_{-\infty}^\infty=\int_{-1}^1+\int_{-\infty}^{-1}+\int_1^\infty$, and use \eqref{JA} to obtain
\begin{eqnarray}
&&\nonumber\pi I(k)=-\frac{g}{4\sqrt2}\int_0^\infty dt(\int_{-\infty}^{-1}+\int_1^\infty) du\Bigg(\cos(ut)\frac{\cosh(g\pi(u+\tfrac{k}{2g}))}{\cosh(2g\pi(u+\tfrac{k}{2g}))}(\Gamma_-(t)-\Gamma_+(t))\\
&&\nonumber\hspace{7cm}+\sin(ut)\frac{\sinh(g\pi(u+\tfrac{k}{2g}))}{\cosh(2g\pi(u+\tfrac{k}{2g}))}(\Gamma_-(t)+\Gamma_+(t))\Bigg)\\
&&\nonumber-\frac{g}{4\sqrt2}\int_{-1}^1du\frac{\cosh(g\pi(u+\tfrac{k}{2g}))}{\cosh(2g\pi(u+\frac{k}{2g}))}2(\log M+\gamma_E+2\log 2)\\
&&\nonumber-\frac{g}{4\sqrt2}\int_{-1}^1du\frac{\cosh(g\pi(u+\tfrac{k}{2g}))}{\cosh(2g\pi(u+\frac{k}{2g}))}\left(\int_0^\infty  dt\left(L\frac{J_0(2gt)}{e^t-1}-\frac{e^{t/2}}{e^t-1}\hat{\rho}_h(t)\right)\cos(2gtu)-2\frac{e^{t/2}}{e^t-1}\right)\\
&&-\frac{g}{4\sqrt2}\int_{-1}^1du\frac{\sinh(g\pi(u+\tfrac{k}{2g}))}{\cosh(2g\pi(u+\frac{k}{2g}))}\int_0^\infty  dt\left(L\frac{J_0(2gt)}{e^t-1}-\frac{e^{t/2}}{e^t-1}\hat{\rho}_h(t)\right)\sin(2gtu)
\end{eqnarray}
At large $g$ we have
\begin{eqnarray}
\nonumber\lefteqn{\int_{-1}^1\left(\frac{\cosh(g\pi(u+\tfrac{k}{2g}))}{\cosh(2g\pi(u+\frac{k}{2g}))}\cos(2gtu)+\frac{\sinh(g\pi(u+\tfrac{k}{2g}))}{\cosh(2g\pi(u+\frac{k}{2g}))}\sin(2gtu)\right)=\frac{2}{\sqrt{2}g}\cos(kt)\frac{e^{t/2}}{e^{t}+e^{-t}}}\\
\nonumber&-&\int_{-\infty}^{-1}du\,e^{g\pi u+\pi k/2}(\cos(2gtu)+\sin(2gtu))-\int_1^{\infty}du\,e^{-g\pi u-\pi k/2}(\cos(2gtu)-\sin(2gtu))\\
&=&\nonumber\frac{\sqrt2}{g}\cos(kt)\frac{e^{t/2}}{e^{t}+e^{-t}}+2\cosh(\tfrac{\pi k}{2})\int_1^\infty du e^{-\pi g u}(\cos(2gtu)-\sin(2gtu))\\
&=&\frac{\sqrt2}{g}\cos(kt)\frac{e^{t/2}}{e^{t}+e^{-t}}-\frac{\sqrt{2}}{ g}e^{-\pi g}\cosh(\tfrac{\pi k}{2})\mbox{Re}\left[\frac{i}{t+i\pi/2}e^{2igt-\pi i/4}\right].
\end{eqnarray}
Similarly we evaluate
\begin{eqnarray}
&&\nonumber(\int_{-\infty}^{-1}+\int_1^\infty) du\Bigg(\cos(ut)\frac{\cosh(g\pi(u+\tfrac{k}{2g}))}{\cosh(2g\pi(u+\tfrac{k}{2g}))}(\Gamma_-(t)-\Gamma_+(t))\\
&&\nonumber\hspace{7cm}+\sin(ut)\frac{\sinh(g\pi(u+\tfrac{k}{2g}))}{\cosh(2g\pi(u+\tfrac{k}{2g}))}(\Gamma_-(t)+\Gamma_+(t))\Bigg)\\
&&\nonumber=2\cosh(\tfrac{\pi k}{2})\int_1^\infty du e^{-\pi g u}\left(\cos(ut)(\Gamma_-(t)-\Gamma_+(t))+\sin(ut)(\Gamma_-(t)+\Gamma_+(t))\right)\\
&&=2\sqrt2\cosh(\tfrac{\pi k}{2})e^{-\pi g} \,\mbox{Re}\left[\frac{1}{t+i\pi g}(\Gamma_+(t)+i\Gamma_-(t))e^{it-i\pi/4}\right].
\end{eqnarray}
Combining the above results we arrive at \eqref{integralres}.

\section{The anomalous dimension}\label{Details2}
The anomalous dimension is given by \eqref{AD1}.
Equivalently, with the help of \eqref{gammasplit} and \eqref{Neumann}, we can write $\gamma=16g^2\gamma_1$. $\gamma_1$ can be obtained by making use of the Wronskian relations, as in \cite{Basso:2008tx, Freyhult:2009my}, which are derived as follows. Starting from \eqref{strongcouplingform} we introduce a new set of equations which contain an auxiliary variable $x$,
\begin{eqnarray}\label{strongcouplingform2}
&&\nonumber\int_0^\infty \frac{dt}{t}J_{2n}(2gt)\left(\frac{\gamma_+(2gt,x)}{1-e^{-t}}-\frac{\gamma_-(2gt,x)}{e^t-1}\right)=xh_{2n}\\
&&\int_0^\infty \frac{dt}{t}J_{2n-1}(2gt)\left(\frac{\gamma_+(2gt,x)}{e^t-1}+\frac{\gamma_-(2gt,x)}{1-e^{-t}}\right)=\frac{1}{2}(1-x)\delta_{n,1}+xh_{2n-1},
\end{eqnarray}
where $x=0$ corresponds to the BES equation \cite{Eden:2006rx,Beisert:2006ez,Basso:2007wd} and $x=1$ corresponds to the new piece added here. Multiplying the equations by $2\cdot2n\gamma_{2n}(x')$ and $2(2n-1)\gamma_{2n-1}(x')$, respectively, summing over $n$ and finally subtracting the two resulting equations we find
\begin{eqnarray}
&&\hspace{-0.7cm}\nonumber\int_0^\infty \frac{dt}{t}\left(\frac{\gamma_+(2gt,x')\gamma_+(2gt,x)-\gamma_-(2gt,x)\gamma_-(2gt,x')}{1-e^{-t}}\right.\\
&&\hspace{1cm}\left.\nonumber+\frac{\gamma_+(2gt,x')\gamma_-(2gt,x)+\gamma_-(2gt,x')\gamma_+(2gt,x')}{e^{t}-1}\right)\\
&&\hspace{-0.7cm}=x\sum_{n=1}^\infty h_{2n}2\,2n\gamma_{2n}(x')-(1-x)\gamma_1(x')-x\sum_{n=1}^\infty h_{2n-1}2(2n-1)\gamma_{2n-1}(x').
\end{eqnarray}
Using that the left hand side is odd under $x\leftrightarrow x'$ and setting $x=1$ and $x'=0$ we find
\begin{eqnarray}
\nonumber\gamma_1&=&\sum_{n=1}^\infty 2\,2nh_{2n}\gamma_{2n}(0)-\sum_{n=1}^\infty 2\,(2n-1)h_{2n-1}\gamma_{2n-1}(0)\\
&=&\nonumber\int_0^\infty dt\left(\left(\frac{L}{2}\frac{J_0(2gt)}{e^t-1}-\frac{L-2}{2}\frac{e^{t/2}}{e^t-1}\right)\frac{\gamma_+^{(0)}(2gt)-\gamma_-^{(0)}(2gt)}{2gt}-2\gamma_1^{(0)}\frac{e^{t/2}}{e^t-1}\right)\\
&-&\int_0^\infty dt\frac{\gamma_+^{(0)}(2gt)-\gamma_-^{(0)}(2gt)}{2gt}\frac{1}{2}\frac{e^{t/2}}{e^t-1}\left(\hat{\rho}_h(t)-\hat{\rho}_h(0)\right),
\end{eqnarray}
where $\gamma_\pm^{(0)}$ denote the solutions to the BES equation.
To obtain the above we used the normalisation condition for the hole density \eqref{norm}. Using $J_0(2gt)=1-2\sum_{n=1}^\infty J_{2n}(2gt)$ and the orthogonality of Bessel functions the first part can be identified with the virtual scaling function $B_L(g)$,
\begin{equation}
\gamma_1=\frac{1}{16g^2}B_L(g)-\int_0^\infty dt\frac{\gamma_+^{(0)}(2gt)-\gamma_-^{(0)}(2gt)}{2gt}\frac{1}{2}\frac{e^{t/2}}{e^t-1}\left(\hat{\rho}_h(t)-\hat{\rho}_h(0)\right).
\end{equation}
Following \cite{Basso:2008tx} we find to leading orders in the large $g$ exansion
\begin{equation}
\gamma_1=\frac{1}{16g^2}B_L(g)+\frac{m_{O(6)}}{8g^2}\int_{-a}^a dk\rho_h(k)\left(\cosh(\tfrac{\pi k}{2})-1\right).
\end{equation}
Putting everything together we obtain \eqref{AD}.

\section{The mass parameter}
The first non-perturbative corrections to the cusp anomalous dimension were computed in \cite{Basso:2009gh}. They are given by the functions $\delta\gamma_\pm(t)$ which satisfy
\begin{eqnarray}\label{deltagammaeqs}
&&\nonumber\int_0^\infty\frac{dt}{t}J_{2n}(t)\left(\frac{\delta\gamma_+(t)}{1-e^{-t/2g}}-\frac{\delta\gamma_-(t)}{e^{t/2g}-1}\right)=0\\
&&\int_0^\infty\frac{dt}{t}J_{2n-1}(t)\left(\frac{\delta\gamma_-(t)}{1-e^{-t/2g}}+\frac{\delta\gamma_+(t)}{e^{t/2g}-1}\right)=0.
\end{eqnarray}
These functions are related to $\hat{\gamma}_\pm(t)$ \eqref{nonpert} as in \cite{Basso:2009gh}
\begin{eqnarray}\label{deltagamma/gammahat}
\delta\gamma_+(t)&=&\hat{\gamma}_+(t)-\frac{2m'}{\pi}\frac{t^2}{t^2+\pi^2g^2}\\
\delta\gamma_-(t)&=&\hat{\gamma}_-(t)+\frac{2gm't}{t^2+\pi^2g^2}
\end{eqnarray}
where they were explicitly written down. The explicit solution reads
\begin{equation}\label{changeofvar2}
\delta\gamma_+(it)+i\delta\gamma_-(it)=\frac{\sin(\tfrac{t}{4g})}{\sqrt{2}\sin(\tfrac{t}{4g}+\tfrac{\pi}{4})}(\delta\Gamma_+(it)+i\delta\Gamma_-(it))
\end{equation}
where
\begin{equation}\label{exactsol}
\delta\Gamma_+(4\pi git)+i\delta\Gamma_-(4\pi git)=\delta f_0(4\pi g t)V_0(4\pi gt)+\delta f_1(4\pi g t)V_1(4\pi gt)
\end{equation}
with
\begin{eqnarray}
\delta f_0(4\pi gt)&=&\Lambda^2\left(\frac{1}{4\pi g}\left(\frac{\Gamma(3/4)\Gamma(1-t)}{2\Gamma(3/4-t)}-\frac{\Gamma(5/4)\Gamma(1+t)}{2\Gamma(5/4+t)}\right)+\mathcal{O}(1/g^2)\right)\\
\delta f_1(4\pi gt)&=&\Lambda^2\left(\frac{1}{4\pi g}\frac{\Gamma(5/4)\Gamma(1+t)}{\Gamma(5/4+t)}+\mathcal{O}(1/g^2)\right)\\
V_n(t)&=&\frac{\sqrt{2}}{\pi}\int_{-1}^1du\left(\frac{1+u}{1-u}\right)^{1/4}\frac{e^{ut}}{(1+u)^n}.
\end{eqnarray}
$\Lambda$ is the nonperturbative scale, $\Lambda^2=-\pi\sqrt{2}gm'm_{O(6)}$.

We will make use of this solution by using the Wronskian relations \eqref{Wronskianresult} to write an expression for the mass parameter, appearing in our effective equations, in terms of $\delta\gamma_\pm(t)$. 
The left hand side of \eqref{Wronskianresult} reads, after the change of variables \eqref{changeofvar2},
\begin{eqnarray}
\nonumber\hspace{-0.5cm} \lefteqn{-\frac{m'}{\pi}\int_0^\infty dt\left[\frac{\pi g}{t^2+(\pi g)^2}\left(\Gamma_-(t)-\Gamma_+(t)\right)+\frac{t}{t^2+(\pi g)^2}\left(\Gamma_-(t)+\Gamma_+(t)\right)\right]}\\
\hspace{-0.5cm}&=&-\frac{m'}{\pi}\int_0^\infty du e^{-\pi g u}\int_0^\infty dt\big[\cos(ut)\left(\Gamma_-(t)-\Gamma_+(t)\right)+\sin(ut)\left(\Gamma_-(t)+\Gamma_+(t)\right)\big].
\end{eqnarray}
Splitting the integral over $u$ as $\int_{0}^\infty=\int_0^1+\int_1^\infty$ and using \eqref{JA} this can be approximated at strong coupling by
\begin{eqnarray}
\nonumber&&\hspace{-0.8cm}{-\frac{m'}{\pi}\int_0^1 due^{-\pi g u}\int_0^\infty dt\left[\left(L\frac{J_0(2gt)}{e^t-1}-\frac{e^{t/2}}{e^t-1}\hat{\rho}_h(t)\right)(\cos(2gtu)+\sin(2gtu))-2\frac{e^{t/2}}{e^t-1}\right]}\\
\nonumber&-&2(\log M+\gamma_E+2\log 2)\frac{m'}{\pi}\int_0^1 due^{-\pi g u}\\
&-&\frac{m_{O(6)}}{\pi}\int_1^\infty du e^{-\pi g u}\int_0^\infty dt\big[\cos(ut)\left(\Gamma_-(t)-\Gamma_+(t)\right)+\sin(ut)\left(\Gamma_-(t)+\Gamma_+(t)\right)\big].
\end{eqnarray}
Integration over $u$ results in 
\begin{eqnarray}
&&\int_0^1 due^{-\pi g u}\left(\cos(2gtu)+\sin(2gtu)\right)=\frac{1}{\sqrt{2}g}\mbox{Re}\left(\frac{-ie^{-i\pi/4}}{t+i\pi /2}(e^{2igt-\pi g}-1)\right)\\
&&\nonumber\int_1^\infty du e^{-\pi g u}\left(\cos(ut)(\Gamma_-(t)-\Gamma_+(t))+\sin(ut)(\Gamma_-(t)+\Gamma_+(t))\right)\\
&&=\sqrt{2}e^{-\pi g}\mbox{Re}\left(\frac{e^{i(t-\pi/4)}}{t+i\pi g}(\Gamma_+(t)+i\Gamma_-(t))\right).
\end{eqnarray}
Hence the left hand side becomes
\begin{eqnarray}
&&\hspace{-0.1cm}\nonumber-\frac{2m'}{\pi^2g}(1-e^{-\pi g})(\log M+\gamma_E+2\log 2)\\
&&\hspace{-0.1cm}\nonumber-\frac{m'}{\pi g}\int_0^\infty\hspace{-0.1cm} dt\left(\hspace{-0.1cm}\left(L\frac{J_0(2gt)}{e^t-1}-\frac{e^{t/2}}{e^t-1}\hat{\rho}_h(t)\right)\hspace{-0.05cm}\mbox{Re}\hspace{-0.05cm}\left(\frac{\tfrac{i}{\sqrt2}e^{-i\pi/4}}{t+i\pi /2}(1-e^{2igt-\pi g})\right)-\frac{2}{\pi }(1-e^{-\pi g})\frac{e^{t/2}}{e^t-1}\right)\\
&&\hspace{-0.1cm}-\frac{m'}{\pi}\int_0^\infty \sqrt{2}e^{-\pi g}\,\mbox{Re}\left(\frac{e^{i(t-\pi/4)}}{t+i\pi g}(\Gamma_+(t)+i\Gamma_-(t))\right)
\end{eqnarray}
For the right hand side of the equation \eqref{Wronskianresult} we find
\begin{eqnarray}
&&\nonumber\hat{\gamma}_1(\log M+\gamma_E+2\log 2)\\
&&+\frac{1}{2}\int_0^\infty dt\left[\left(L\frac{J_0(2gt)}{e^t-1}-\frac{e^{t/2}}{e^t-1}\hat{\rho}_h(t)\right)\frac{\hat{\gamma}_-(2gt)-\hat{\gamma}_+(2gt)}{4gt}-2\hat{\gamma_1}\frac{e^{t/2}}{e^t-1}\right].
\end{eqnarray}
With the help of \eqref{deltagamma/gammahat} and the explicit expression for $\hat{\gamma}_1$,
\begin{equation}
\hat{\gamma}_1=-\frac{2m'}{\pi^2g}+\frac{m'm_{O(6)}}{4\sqrt{2}},
\end{equation}
 the right hand side is rewritten and \eqref{Wronskianresult} can be rearranged as in \eqref{Wronskianlargeg}. This relation allows us to write the mass parameter as \eqref{mass2}.

Using the relation for the Bessel functions, $J_0(2gt)=1-2\sum_{n=1}^\infty J_{2n}(2gt)$ and the equations \eqref{deltagammaeqs} the mass parameter \eqref{mass2} can be further rewritten as
\begin{eqnarray}
m&=&\nonumber m_{O(6)}\left(\log M+\gamma_E+2\log 2\right)+\frac{L}{\pi\sqrt{g}}e^{-\pi g}\\
&+&\nonumber\frac{4\sqrt2 g}{m'}\int_0^\infty\left(\frac{\delta \gamma_-(2gt)}{gt(e^t-1)}-\frac{\delta\gamma_+(2gt)}{gt(1-e^{-t})}-\frac{m_{O(6)}m'}{4\sqrt2g}\frac{e^{t/2}}{e^t-1}\right)\\
&-&\nonumber\frac{(L-2)\sqrt2g}{m'}\int_0^\infty dt\left(\frac{\delta\gamma_-(2gt)}{gt(e^{t/2}+1)}+\frac{\delta\gamma_+(2gt)}{gt(e^{-t/2}+1)}\right)\\
&-&\frac{4\sqrt{2}g}{m'}\int_0^\infty dt\frac{e^{t/2}}{e^t-1}\left(\hat{\rho}_h(t)-\hat{\rho}_h(0)\right)\frac{\delta\gamma_-(2gt)-\delta\gamma_+(2gt)}{2gt}.
\end{eqnarray}
After the change of variables \eqref{changeofvar2} the above expression reads
\begin{eqnarray}\label{mass3}
m&=&\nonumber m_{O(6)}\left(\log M+\gamma_E+2\log 2\right)+\frac{L}{\pi\sqrt{g}}e^{-\pi g}\\
&-&\nonumber\frac{4\sqrt2 g}{m'}\int_0^\infty\left(\frac{\delta\Gamma_-(2gt)+\delta\Gamma_+(2gt)}{2gt}+\frac{m_{O(6)}m'}{4\sqrt2g}\frac{e^{t/2}}{e^t-1}\right)\\
&-&\nonumber\frac{(L-2)\sqrt2g}{2m'}\int_0^\infty \frac{dt}{t}\left[\left(1-\frac{\cosh(\tfrac{t}{4g})}{\cosh(\tfrac{t}{2g})}\right)(\delta\Gamma_-(t)+\delta\Gamma_+(t))+\frac{\sin(\tfrac{t}{4g})}{\cosh(\tfrac{t}{2g})}(\delta\Gamma_-(t)-\delta\Gamma_+(t))\right]\\
&-&\frac{4\sqrt{2}g}{m'}\int_0^\infty dt\frac{e^{t/2}}{e^t-1}\left(\hat{\rho}_h(t)-\hat{\rho}_h(0)\right)\frac{\delta\gamma_-(2gt)-\delta\gamma_+(2gt)}{2gt}.
\end{eqnarray}
Of the integrals in this expression we first consider
\begin{eqnarray}\label{massintegral}
\int_0^\infty\left(\frac{\delta\Gamma_-(2gt)+\delta\Gamma_+(2gt)}{2gt}+\frac{m_{O(6)}m'}{4\sqrt2g}\frac{e^{t/2}}{e^t-1}\right).
\end{eqnarray}
The functions special functions $V_{0,1}(t)$ in \eqref{exactsol} can be written in terms of Bessel functions,
\begin{eqnarray}
V_0(t)&=&\frac{1}{2}\sum_{k=0}^\infty (-1)^{k+1}\frac{\Gamma(k-1/2)}{\Gamma(k+1)\Gamma(1/2)}\left(J_{2k}(it)+iJ_{2k-1}(it)\right)\nonumber\\
V_1(t)&=&2\sum_{k=0}^\infty (-1)^{k+1}\frac{\Gamma(k-1/2)}{\Gamma(k+1)\Gamma(1/2)}\left(-(k-1/2)J_{2k}(it)+ikJ_{2k-1}(it)\right),
\end{eqnarray}
With this we find that the integral \eqref{massintegral} reduces, at strong coupling, to
\begin{eqnarray}
&-&\nonumber\frac{m'm_{O(6)}}{4\sqrt2g}\int_0^\infty dt\left(\frac{J_0(2gt)}{t}-\frac{e^{t/2}}{e^t-1}\right)\\
&-&\nonumber\frac{m'm_{O(6)}}{4\sqrt2g}\sum_{k=1}^\infty(-1)^{k+1}\frac{\Gamma(k-1/2)}{\Gamma(k+1)\Gamma(1/2)}\left(\frac{1/2-k}{2k}+\frac{k}{2k-1}\right)\\
&=&-\frac{m'm_{O(6)}}{4\sqrt2g}(-\log g-\gamma_E-\log 2)
\end{eqnarray}
The further integrals will be subleading in the limit \eqref{O(6)limit} and hence we find the result \eqref{mass5} and \eqref{mass4}.

\end{document}